\input harvmac


\def\np#1#2#3{Nucl. Phys. {\bf B#1} (#2) #3}
\def\pl#1#2#3{Phys. Lett. {\bf #1B} (#2) #3}
\def\prl#1#2#3{Phys. Rev. Lett. {\bf #1} (#2) #3}
\def\prd#1#2#3{Phys. Rev. {\bf D#1} (#2) #3}

\def\jhep#1#2#3{JHEP {\bf#1}(#2) #3}
\def\jmp#1#2#3{J. Math Phys. {\bf #1} (#2) #3}

\def\ijmp#1#2#3{Int.~J.~Mod.~Phys. {\bf #1} (#2) #3}
\def\atmp#1#2#3{Adv.~Theor.~Math.~Phys.{\bf #1} (#2) #3}

\def\IB{\relax\hbox{$\inbar\kern-.3em{\rm B}$}}
\def\IC{\relax\hbox{$\inbar\kern-.3em{\rm C}$}}
\def\ID{\relax\hbox{$\inbar\kern-.3em{\rm D}$}}
\def\IE{\relax\hbox{$\inbar\kern-.3em{\rm E}$}}
\def\IF{\relax\hbox{$\inbar\kern-.3em{\rm F}$}}
\def\IG{\relax\hbox{$\inbar\kern-.3em{\rm G}$}}
\def\IGa{\relax\hbox{${\rm I}\kern-.18em\Gamma$}}
\def\IH{\relax{\rm I\kern-.18em H}}
\def\IK{\relax{\rm I\kern-.18em K}}
\def\IL{\relax{\rm I\kern-.18em L}}
\def\IP{\relax{\rm I\kern-.18em P}}
\def\IR{\relax{\rm I\kern-.18em R}}
\def\IZ{\relax\ifmmode\mathchoice{
\hbox{\cmss Z\kern-.4em Z}}
{\hbox{\cmss Z\kern-.4em Z}}
{\lower.9pt\hbox{\cmsss Z\kern-.4em Z}}
{\lower1.2pt\hbox{\cmsss Z\kern-.4em Z}}
\else{\cmss Z\kern-.4em Z}\fi}
\def\II{\relax{\rm I\kern-.18em I}}

\def\ttb{Type $\II$B string theory}
\def\ndt{{\noindent}}

\def\CN{{\cal N}}

\def\CV{{\cal V}}

\def\p{\partial}


\def\kb{\bar{k}}
\def\lb{\bar{l}}
\def\mb{\bar{m}}
\def\nb{\bar{n}}

\def\p{\partial}

\def\Tr{{\rm Tr}}


\def\inbar{\,\vrule height1.5ex width.4pt depth0pt}

\font\cmss=cmss10 \font\cmsss=cmss10 at 7pt

\def\a{{\alpha}}
\def\b{{\beta}}

\def\g{{\gamma}}

\def\ve{{\varepsilon}}
\def\m{{\mu}}
\def\n{{\nu}}
\def\l{{\lambda}}
\def\s{{\sigma}}



\lref\tasi{J.~Polchinski, ``TASI Lectures on D-Branes'',
  {\tt hep-th/9611050}}

\lref\dgmorb{M.R. Douglas, B.R. Greene and D.R. Morrison,
``Orbifold Resolution by D-branes'', \np{505}{1997}{84},
  {\tt hep-th/9704151}}

\lref\dougegs{M.R. Douglas, ``Enhanced Gauge Symmetry
in M(atrix) theory'', \jhep{007}{1997}{004},
  {\tt hep-th/9612126}}

\lref\jmorb{C.V. Johnson and R.C. Myers, ``Aspects of
Type $\II$B Theory on ALE Spaces'', \prd{55}{1997}{6382},
  {\tt hep-th/9610140}}

\lref\gipol{E.G.~Gimon and J.~Polchinski, ``Consistency Conditions for
Orientifolds and D Manifolds'', \prd{54}{1996}{1667},
  {\tt hep-th/9601038}}

\lref\polten{J.~Polchinski, ``Tensors from K3 Orientifolds'',
\prd{55}{1997}{6423}, {\tt hep-th/9606165}}

\lref\kmvgeom{S.~Katz, P.~Mayr and C.~Vafa,
  ``Mirror Symmetry and
  Exact Solution of 4D $\CN=2$ Gauge Theories -- I'',
  {\tt hep-th/9706110}}

\lref\branegeom{H.~Ooguri and C.~Vafa, ``Geometry of $\CN=1$ dualities
in four dimensions'', \np{500}{1997}{62}; {\tt hep-th/9702180}}

\lref\absorb{S. Gubser and I. Klebanov,``Absorption by Branes
and Schwinger Terms in the World Volume Theory,''
\pl{413}{1997}{41}}

\lref\mfour{E.~Witten, ``Solutions of Four-dimensional
  Field Theories via M-theory'', \np{500}{1997}{3}; {\tt hep-th/9703166}}

\lref\aspegs{P.S. Aspinwall, ``Enhanced Gauge Symmetries and K3
Surfaces'', \pl{357}{1995}{329}, {\tt hep-th/9507012}}

\lref\bsvegs{M. Bershadsky, V. Sadov and C. Vafa, ``D-strings on
D-manifolds'',
\np{463}{1996}{398}}

\lref\dixon{L. Dixon, D. Friedan, E. Martinec and S. Shenker,
``The Conformal Field Theory of Orbifolds,''
\np{282}{1987}{13}}

\lref\br{M. Bershadsky and A. Radul,
``Conformal Field Theories with Additional $\IZ_N$ Symmetry,''
\ijmp{A2}{1987}{165}}

\lref\kwdiscrete{L.M. Krauss and F. Wilczek,
  ``Discrete Gauge Symmetries
in Continuum Theories'', \prl{62}{1989}{1221}}

\lref\reidrev{M. Reid, ``McKay correspondence'',
  {\tt alg-geom/9702016}}

\lref\suthree{W.M.~Fairbanks, T.~Fulton and W.H.~Klink, ``Finite and
Disconnected Subgroups of $SU(3)$ and their Application to the
Elementary Particle Spectrum'', \jmp{5}{1964}{1038}}
\lref\sosix{W.~Plesken and M.~Pohst, Math. Comp. {\bf 31} (1977) 552}
\lref\bkv{M.~Bershadsky, Z.~Kakushadze and  C.~Vafa, 
``String expansion as large N expansion of gauge theories",
\np {523} {1998} {59}, {\tt hep-th/9803076}}
\lref\bj{M.~Bershadsky and A.~Johansen, {\tt hep-th/9803249}}

\lref\KT{I.~R.~Klebanov and A.~A.~Tseytlin, ``D-Branes and
Dual Gauge Theories in Type 0 Strings,''
\np{546}{1999}{155}, {\tt hep-th/9811035}}

\lref\KTc{I.~R.~Klebanov and A.~A.~Tseytlin,
  ``Non-supersymmemtric CFT from Type 0 String Theory,''
\jhep{9903}{1999}{015}, {\tt hep-th/9901101}\semi
 I.~R.~Klebanov,
``Tachyon Stabilization in the AdS/CFT Correspondence,''
{\tt hep-th/9906220}}

\lref\DM{M.~Douglas and G.~Moore,
``D-branes, quivers, and ALE instantons,'' {\tt hep-th/9603167}}

\lref\DH {L.~Dixon and J.~Harvey,
``String theories in ten dimensions without
space-time supersymmetry", \np{274}{1986}{93} \semi
N.~Seiberg and E.~Witten,
``Spin structures in string theory", \np{276}{1986}{272}\semi
C.~Thorn, unpublished}

\lref\berg{O.~Bergman and M.~Gaberdiel, ``A Non-supersymmetric Open
String Theory and S-Duality,'' \np{499}{1997}{183},
{\tt hep-th/9701137}}

\lref\jthroat{J.~Maldacena,
``The Large N limit of superconformal field theories and
  supergravity,'' \atmp{2}{1998}{231},
{\tt  hep-th/9711200} }

\lref\gkp{S.S.~Gubser, I.R. Klebanov, and A.M. Polyakov,
  ``Gauge theory correlators from noncritical string theory,''
  \pl{428}{1998}{105}, {\tt hep-th/9802109}}

\lref\EW{E.~Witten, ``Anti-de Sitter space and holography,''
\atmp{2}{1998}{253},
{\tt hep-th/9802150}}

\lref\AP{A.M.~Polyakov, ``The Wall of the Cave,''
{\tt hep-th/9809057}}

\lref\KS{S.~Kachru and E.~Silverstein, ``4d conformal field theories
and strings on orbifolds,''  \prl{80}{1998}{4855},
{{\tt hep-th/9802183}}.}
\lref\LNV{A.~Lawrence, N.~Nekrasov and C.~Vafa, ``On conformal field
theories in four dimensions,'' \np{533}{1998}{199},
{{\tt hep-th/9803015}}}

\lref\NS{N. Nekrasov and S. Shatashvili,
``On non-supersymmetric CFT in four dimensions,''
{\tt hep-th/9902110.}, L.~Okun Festschrift, North-Holland,
in press}

\lref\JM{J. Minahan, ``Glueball Mass Spectra and Other Issues for
Supergravity Duals of QCD Models,'' {\tt hep-th/9811156}}

\lref\JMnew {J. Minahan, ``Asymptotic Freedom and Confinement from Type 0
String Theory,'' {\tt hep-th/9902074}}

\lref\KTnew{I.R. Klebanov and A.A. Tseytlin, ``Asymptotic Freedom and
Infrared Behavior in the Type 0 String Approach to Gauge Theory,''
\np{547}{1999}{143}, {\tt hep-th/9812089} }
\lref\KHrev{I.~R.~Klebanov and A.~Hashimoto, 
``Scattering of Strings from D-branes,'' {\tt  hep-th/9611214}}

\lref\Blum{R. Blumenhagen, A. Font and D. Lust,
``Non-Supersymmetric Gauge Theories from
D-Branes in Type 0 String Theory,''
{\tt hep-th/9906101.}}

\lref\BCR{M. Bill\' o, B. Craps and  F. Roose, ``On D-branes in
Type 0 String Theory,'' {\tt hep-th/9902196.}
}

\lref\Zar{K. Zarembo, ``Coleman-Weinberg Mechanism and Interaction of
D3-branes in Type 0 String Theory, {\tt hep-th/9901106}}

\Title{\vbox
{\baselineskip 10pt
\hbox{PUPT-1891}
\hbox{ITEP-TH-48/99}
\hbox{NSF-ITP-99-27}
\hbox{YCTP-P24-99}
\hbox{hep-th/9909109}
{\hbox{   }}}}
{\vbox{\vskip -30 true pt
\centerline {An Orbifold of Type 0B Strings}
\medskip
\centerline {and Non-supersymmetric Gauge Theories}
\medskip
\vskip4pt }}
\vskip -20 true pt
\centerline{ Igor R.~Klebanov,$^{1}$ Nikita A.~Nekrasov$^{2}$ and Samson
 L.~Shatashvili$^{3}$\footnote{$^{\dagger}$}{On leave of absence
from Steklov Mathematical Institute, St.Petersburg, Russia} }
\smallskip\smallskip
\centerline{$^{1,2}$ \it Joseph Henry
Laboratories, Princeton University, Princeton, New Jersey 08544}
\centerline{$^{2}$ \it Institute for Theoretical and Experimental
Physics, 117259 Moscow, Russia}
\centerline {$^{3}$ \it Department of Physics, Yale University, New Haven, CT
06520 }

\bigskip\bigskip
\centerline {\bf Abstract}
\baselineskip12pt
\noindent
\medskip

We study a
${{\IZ}}_2$ orbifold of Type 0B string theory by reflection of
six of the coordinates (this theory may also be thought of as a
${{\IZ}}_4$ orbifold of {\ttb} by a rotation by $\pi$ in three independent
planes). We show that the only massless mode localized on the fixed
fourplane $\IR^{3,1}$ 
is a $U(1)$ gauge field. After introducing D3-branes parallel
to the fixed fourplane we find non-supersymmetric non-abelian
gauge theories on their worldvolume. One of our results is that
the theory on
equal numbers of electric and magnetic D3-branes placed at the
fourplane is the ${{\IZ}}_4$ orbifold of
${\cal N}=4$ supersymmetric
Yang-Mills theory by the center of its R-symmetry group.

\bigskip

\Date{09/99}

\noblackbox \baselineskip 15pt plus 2pt minus 2pt

\newsec{Introduction}

The discovery of AdS/CFT correspondence
\refs{\jthroat,\gkp,\EW}
has led to new insights into four 
dimensional gauge theories. In particular the results on
D3-branes near
orbifold singularities \DM\ naturally lead to quotients of
the basic
${\cal N}=4$ duality by discrete subgroups of the $SU(4)$
R-symmetry \refs{\KS,\LNV}.
A closely related development has been the study of
 gauge theories on
D3-branes of Type 0B theory, which is an NSR string with
the non-chiral
GSO projection $(-1)^{F+\tilde F}=1$
which breaks all spacetime supersymmetry \DH. Type
0B string can also be studied in the GS framework,
as we show below.

The Type 0A and 0B models have twice as many massless
RR fields as their Type $\II$ cousins, hence they
also possess twice as many D-branes \KT. For example,
since Type 0B
spectrum has an unrestricted 4-form gauge potential, there are
two types of D3-branes: those that couple electrically to this
gauge potential, and those that couple magnetically.
Very importantly, the weakly coupled spectrum of open strings on
Type 0 D-branes does not contain tachyons after the GSO
projection $(-1)^{F_{open}}=1$
is implemented \refs{\KT,\AP,\berg}. Thus, gauge
theories living on
such D-branes do not have obvious instabilities
at weak coupling.
This suggests via
the gauge field/string duality that the bulk
tachyon instability
of Type 0 theory may be cured as well \refs{\KT,\KTc}.

The simplest example of AdS/CFT duality in a Type 0 context
follows from considering $N$ electric D3-branes coincident with
$N$ magnetic ones \KTc.
For such a stack the net tachyon tadpole cancels so that there
exists a classical solution with $T=0$. In fact, since the stack couples
to the selfdual part of the 5-form field strength, the Type 0B
3-brane classical
solution is identical to the Type $\II$B one. Taking the throat
limit suggests that the low-energy field theory on
$N$ electric and $N$ magnetic D3-branes is dual to the $AdS_5\times S^5$
background of Type 0B theory and is therefore conformal in the planar
limit \KTc. This theory is the $U(N)\times U(N)$
gauge theory coupled to six adjoint scalars of the first $U(N)$,
six adjoint scalars of the second $U(N)$, and Weyl fermions in the
bifundamental representations -- four in the
$({\bf N}, \overline {\bf N})$ and four in the
$(\overline {\bf N}, {\bf N})$ (the $U(1)$ factors decouple in
the infrared). This theory is a
${\IZ}_2$ projection of the ${\cal N}=4$ $U(2N)$ gauge
theory \KTc.  The ${\IZ}_2$ is generated by
$(-1)^{F_s}$, where $F_s$ is the fermion number,
together with conjugation by
$\pmatrix{ I&  0\cr  0 & -I\cr}$
where $I$ is the $N \times N$ identity matrix.
This is related to the fact that
Type 0 string theories may be viewed
as $(-1)^{F_s}$ orbifolds of the corresponding Type $\II$ theories \DH.
In \NS\ it was pointed out that $(-1)^{F_s}$ is identified with 
an element of the center of the $SU(4)$
R-symmetry, hence this ${\IZ}_2$ projection of the ${\cal N}=4$ theory
belongs to the class studied in \refs{\KS,\LNV}. In particular
this observation explains why the theory is conformal in the large
$N$ limit (this follows from the arguments of \bkv),
and also why the dual string theory background is
$AdS_5 \times S^5$ -- the reason being that the orbifold
group ${{\IZ}}_2$ belongs to the spin cover of the Lorentz
group
and projects to identity in the rotation group. Therefore
the geometry is not affected by the quotienting.

In view of this observation it is of further interest to study the
quotient of the maximally
supersymmetric theory by the full center ${\IZ}_{4}$ of $SU(4)$ \NS. 
This theory has

\item{$\bullet$} the
gauge group $U(N)^4$ coupled to a chiral field content:

\item{$\bullet$}
four quadruples
of bi-fundamental fermions
transforming in $({\bf N}_{\bf i}, {\overline {\bf N}}_{{\bf i}+1})$,
${\bf i} = 0,1,2,3\semi \quad 4 \equiv 0$,

\item{$\bullet$}
four sextets of bi-fundamental
scalars in
$({\bf N}_{\bf i}, {\overline {\bf N}}_{{\bf i}+2})$.

\ndt
Can this gauge theory be embedded into string theory? Since the generator
of ${\IZ}_4$ changes the sign of the six scalar fields in the ${\cal N}=4$
multiplet, a good guess is that this theory occurs on D3-branes of Type
0B theory orbifolded by reflection of six spacetime coordinates \NS.
In this paper we confirm this guess and also derive some further
properties of this ${\IZ}_2$ orbifold of Type 0B theory (more general orbifolds
of Type 0B were studied in \refs{\Blum,\BCR}).
One interesting result is that the only massless mode on the fixed
fourplane $\IR^{3,1}$ is a photon. After D3-branes are added,
this $U(1)$ gauge field
couples to the the theory on the D3-branes, and we comment on
some features of this coupling.
We regard this as an interesting non-supersymmetric analogue of what
happens in orbifolding Type $\II$B theory by reflection of four of the
coordinates.
There 
we find a tensor theory  on the fixed six-plane $\IR^{5,1}$.

\newsec{ The field theories.}
Our goal is to study a certain ${{\IZ}}_{4}$ orbifold of Type $\II$B
string theory.\foot{Notice that there are many
possibilities of embedding the group ${\IZ}_4$ into the
global symmetry group of the \ttb. We consider the embedding
which does not break the global symmetry.} Let us
consider the orbifold of Type $\II$B theory by the
${{\IZ}}_{2}$ reflection
of six out of ten space-time coordinates:
\eqn\refl{
X^{M} \to - X^{M}, \quad M = 4, \ldots, 9
\ .}
Under the corresponding
breaking of   the Lorentz group
$SO(1,9) \to SO(1,3) \times SO(6)$ the ten-dimensional Mayorana-Weyl
spinors
decompose as:
$$
{\bf 16} = {\bf 2}_{\IC}^{SO(1,3)}  \otimes {\bf 4}^{SO(6)}\ .
$$
The transformation \refl\
acts on the space-time spinors via multiplication by $i$
and this ${\IZ}_{2}$ becomes actually ${{\IZ}}_{4} \in SU(4) =
Spin(6)$.

The generator of this ${{\IZ}}_{4}$ is a rotation by
$\pi$ in the
$45$, $67$ and $89$ planes, ${\rm exp}\ \pi i(J_{45} + J_{67} +
J_{89})$. This group has a ${{\IZ}}_{2}$ subgroup generated
by ${\rm exp}\ 2 \pi i (J_{45} + J_{67} +
J_{89})$. If we orbifold the      {\ttb}
by this ${{\IZ}}_{2}$ then we get the
Type 0B string. Indeed, it is known that 0B
is an orbifold of
$\II$B by $(-1)^{F_s}$ where $F_s$ is the spacetime fermion number \DH,
and ${\rm exp}\ 2 \pi i (J_{45} + J_{67} + J_{89})$ acts
on all fields
as $(-1)^{F_s}$. The
subsequent orbifolding of Type 0B string by
${\rm exp}\ \pi i (J_{45} + J_{67} + J_{89})$ is actually a
${{\IZ}}_{2}$ operation because the theory has no
fermions in its
spectrum. Thus, the theory we are studying may also be thought of as
a ${{\IZ}}_{2}$ orbifold of Type 0B string by reflection of six
coordinates \refl.

The same conclusion applies to the field theory on D3-branes.
Take $N$ D3-branes placed at the fixed point of
the discrete group.
The theory induced on the world volume of
these branes will have
four $U(N)$ gauge groups, four quadruples
of bi-fundamental fermions
transforming in $({\bf N}_{\bf i}, {\overline {\bf N}}_{{\bf i}+1})$,
four sextets of
bi-fundamental scalars in $({\bf N}_{\bf i},
{\overline {\bf N}}_{{\bf i}+2})$,
${\bf i} = 0,1,2,3;
{\bf i} \equiv {\bf i} +4$ \NS.

One of our new results is that
this theory can be obtained as a slightly unnatural ${\IZ}_{2}$ orbifold of the
field theory living on the collection of $N$ self-dual threebranes
in Type 0B theory.
In that theory we have two $U(N)$ gauge groups,
two quadruples
of fermions $\psi_{12} \in ({\overline {\bf N}}_{1}, {\bf N}_{2}), \quad
\psi_{21} \in ({\overline {\bf N}}_{2}, {\bf N}_{1})$, and two sextets of
adjoint scalars $\phi_{1}, \phi_{2}$.
This theory has a symmetry:
\eqn\zitwo{
\pmatrix{\phi_{1}\cr \phi_{2}\cr} \mapsto
\pmatrix{- \phi_{1}\cr  - \phi_{2} \cr}, \qquad
\pmatrix{\psi_{12} \cr \psi_{21}}
\mapsto
\pmatrix{- \psi_{12} \cr \psi_{21}}\ .}
Indeed the Yukawa couplings are of the form
$\phi_1 \psi_{12} \psi_{21}$ and
$\phi_2 \psi_{21} \psi_{12}$, hence they
are invariant under \zitwo.
If we add the standard action of ${\IZ}_{2}$ on Chan-Paton indices
(i.e. sending $N \to 2N$ and conjugating by
$\pmatrix{ I&  0\cr  0 & -I\cr}$ in both of the $U(2N)$ groups,
where $I$ is the $N \times N$ identity matrix) then the resulting theory
has the
correct field content and the interactions to be
identified with the ${{\IZ}}_{4}$ orbifold of the ${\cal N}=4$ theory.

\newsec{Closed String Theory}

\subsec{${{\IZ}}_{4}$ orbifold of Type $\II$B in the GS formalism}

Let us consider the {\ttb} in the Green-Schwarz formalism.
In the light cone gauge the action looks like:
\eqn\actn{S = \int d\tau d\sigma \left(
\p_{+} X^{i} \p_{-} X^{i}
- i S^{a} \p_{+} S^{a} - i {\tilde S}^{a} \p_{-} {\tilde S}^{a} \right) }
where the fermions $S^{a}, {\tilde S}^{a}$ transform in ${\bf 8}_{s}$ of
$Spin(8)$, and the scalars $X^{i}$ are in ${\bf 8}_{v}$.

Now let us rewrite this action in the form where the $SU(4)$ invariance is
manifest. To this end we  introduce the complex fermions:
\eqn\bees{b^{m} = {\sqrt{2}} \left( S^{2m} + i S^{2m-1} \right),
\quad {\tilde b}^{m} = {\sqrt{2}} \left( {\tilde S}^{2m} + i {\tilde
S}^{2m-1} \right), \quad m = 1,2,3,4}
The scalars are represented as:
\eqn\sclrs{
X^{i} = \left( X, \, {\bar X}, \, X^{mn} = - X^{nm} \right),
\qquad m,n=1,\ldots,4}
The action \actn\ is rewritten as:
\eqn\actni{S = S (X, \bar X) - i \sum_{m} \int \left(  b^{\mb} \p_{+}
b^{m}
+ {\tilde b}^{\mb} \p_{-} {\tilde b}^{m} \right) +
{1\over 4} \sum \epsilon_{mnkl} \p_{+} X^{kl} \p_{-} X^{mn}
}
We wish to perform the  orbifold projections with respect to the
center of $SU(4)$.

\ndt Naturally we get four sectors.

\ndt{\it Untwisted sector.} In this sector all $X^{i}$'s
and $b^{m}, {\tilde b}^{m}$ are integer moded.
The zero point energy vanishes both in left and right sectors and the
projection of the massless modes gives:
\eqn\proj{\eqalign{SO(8
): \quad & \left( {\bf 8}_{v} \oplus {\bf 8}_{c} \right) \otimes
\left( {\bf 8}_{v} \oplus {\bf 8}_{c} \right) \to\cr
SU(4) : \quad &
{\bf 6} \otimes {\bf 6} \bigoplus 4 \left( {\bf 1} \otimes {\bf 1} \right)
\bigoplus {\bf 4} \otimes {\bf \bar 4}
\bigoplus {\bf\bar 4} \otimes {\bf 4}
\cr}
}
So, all space-time fermions are projected out, while out of bosons
we have got the components of the metric and $B$-field tangent
to the fixed plane ($4 \left({\bf 1} \otimes {\bf 1}\right)$),
orthogonal to the fixed
plane plus the dilaton
($\bf 6 \otimes \bf 6$),
and components of the RR forms which are either transverse or
longitudinal with respect to the fixed plane.

\ndt {\it Twisted sectors}.
$X, \bar X$ are always integer moded. They contribute $-{1\over 12}$ to
the
zero-point energy both on the left and on the right.

\ndt {\it Twists by $\pm i$:}
\eqn\bnd{\eqalign{& X^{mn} ( \sigma + 2\pi ) = - X^{mn} (\sigma)   \cr
& b^{m} (\sigma + 2\pi) = \pm i b^{m} (\sigma) \cr
& {\tilde b}^{m} ( \sigma
+ 2\pi) = \pm i {\tilde b}^{m} ( \sigma) \cr}}

\ndt the zero-point energy of the states
$\vert \pm i \rangle_{L,R}$ is given by
\eqn\zrpttw{
E_{L} = E_{R}  =  - {1\over 12} + 3 ( -{1\over 12} + {1\over 2}
{1\over{2^2}} )
+ 4 ( {1\over 12} - {1\over 2} {3\over{4^2}} ) = 0}
so we get a massless mode for both $+i$ and $-i$ sectors.
Later we will see that these states are not
scalars but, somewhat surprisingly, the positive and negative
helicity components
of a photon in $3+1$ dimensions.

Notice that the next excited levels have $L_{0}={\bar L}_{0}={1\over 4}$
and transform in the rank two tensor representations of
$SU(4)$:
\eqn\nxtlv{\eqalign{
& b^{m}_{- {1\over 4}}
\vert + i \rangle_{L}
\otimes
{\tilde b}^{n}_{ - {1\over 4}}
\vert + i \rangle_{R}
\, \in {\bf 4} \otimes {\bf 4}    \cr
& b^{\mb}_{- {1\over 4}} \vert - i \rangle_{L}
\otimes {\tilde b}^{\nb}_{ - {1\over 4}}
\vert  - i \rangle_{R} \, \in {\bf \bar 4} \otimes
{\bf \bar 4}\cr}}

\ndt {\it Twist by $-1$:}
all bosons are periodic, the fermions are anti-periodic.
The zero point
energy is negative:
\eqn\gsta{E_{L} = E_{R} =
4 ( - {1\over 2} {1\over{2^2}} ) = - {1\over 2}}
so the ground state
$\vert -1 \rangle_{L} \otimes \vert -1 \rangle_{R}$
is
tachyonic.
In the NSR language this is the tachyon from the $(NS-,NS-)$
sector of Type 0B string.
The next level is massless:
\eqn\nxtlvi{
S^{a}_{-\half} \vert -1 \rangle_{L} \otimes
{\tilde S}^{b}_{-\half} \vert -1 \rangle_{R} }
and before the projection gives rise to the extra set of RR
even forms, just like in Type 0B theory.
After the projection it doubles the
set of forms we got in the untwisted sector, i.e.  we get extra
${\bf 4}\otimes {\bf \bar 4} \bigoplus {\bf\bar 4} \otimes {\bf 4}$.

So, in addition to the ${{\IZ}}_2$ projected bulk modes of Type 0B theory
the closed string sector has extra massless modes
propagating along the fixed fourplane. We will identify these
modes with those of a massless vector field.

\subsec{${{\IZ}}_{4}$ orbifold of Type $\II$B in
the NSR formalism}

In the Neveu-Schwarz-Ramond formalism we have only two twisted sectors,
as the space-time part of the orbifold group is only ${\IZ}_{2}$, not
${\IZ}_4$.
Nevertheless, we have effectively the same number of the sectors
due to the two different possibilities of the GSO projection:
$(-1)^F = (-1)^{\tilde F}=1$ and $(-1)^F = (-1)^{\tilde F}=-1$.

It is quite clear that the GS sectors $1$ and $-1$ correspond
to the untwisted sector of NSR formalism. The GS sector $1$
corresponds to NSR states with
$(-1)^F = (-1)^{\tilde F}=1$, i.e. $(NS+,NS+)$ and $(R+,R+)$,
while the GS sector $-1$
corresponds to NSR states with
$(-1)^F = (-1)^{\tilde F}=-1$, i.e. $(NS-,NS-)$ and $(R-,R-)$.
Similarly, the GS sectors $i$ and $-i$ correspond to
the $-1$ twisted sector of NSR with
$(-1)^F = (-1)^{\tilde F}=1$ and $(-1)^F = (-1)^{\tilde F}=-1$
respectively. Let us see how this works in some detail.

\ndt {\it Untwisted sector:}
In the NS sector we have the following boundary conditions:
\eqn\unns{\eqalign{
X^{\mu} ( {\sigma} + 2\pi ) = X^{\mu} ({\sigma}), & \quad
X^{M} (\sigma + 2\pi ) = X^{M} ( \sigma)\cr
\psi^{\mu} ( {\sigma} + 2\pi ) = - \psi^{\mu} (\sigma), & \quad
\psi^{M} (\sigma + 2\pi ) = - \psi^{M} ( \sigma)\cr}}
The ground state has $L_{0} = {\bar L}_{0} = - {1\over 2}$,
which gives the tachyon $T$. This is the state present
in the GS $-1$ sector \gsta. The massless
level is generated by
\eqn\msls{
\psi^{i}_{-\half} \vert 0 \rangle_{L} \otimes
\tilde\psi^{j}_{-\half}
\vert 0 \rangle_{R}}
which is in ${\bf 8}_{v} \otimes {\bf 8}_{v}$ of $SO(8)$ and
then the projection leaves only the purely transverse or purely longitudinal
modes. These are the states we found in the GS untwisted sector \proj.

In the R sector
both the bosons and the fermions are integer moded,
the ground state is supersymmmetric and we have 8 fermion zero modes both
in the left and right sectors. In the $(R+,R+)$ sector, which
corresponds to the GS untwisted sector, we find states in the
${\bf 4} \otimes {\bf\bar 4} \bigoplus {\bf\bar 4} \otimes {\bf 4}$.
The same representation is found in the $(R-,R-)$ sector which
corresponds to the GS $-1$ sector.

\ndt {\it Twisted sector:} the NS boundary conditions are:
\eqn\twns{\eqalign{
X^{\mu} ( {\sigma} + 2\pi ) = X^{\mu} (\sigma), & \quad
X^{M} (\sigma + 2\pi ) = - X^{M} ( \sigma)\cr
\psi^{\mu} ( {\sigma} + 2\pi ) = - \psi^{\mu} (\sigma) , &
\quad
\psi^{M} (\sigma + 2\pi ) = + \psi^{M} ( \sigma)\cr}}
The ground state has zero-point
energy $L_{0} = {\bar L}_{0} = {1\over 4}$, just like
the first excited level in the GS description of the $i$
and $-i$ twisted sectors \nxtlv.
The fermions $\psi^M$ have zero modes so this
excited level is a tensor product of two spinors of $SO(6)$,
in
agreement with what we wrote in the GS section:
in the $(NS+,NS+)$ sector we find
the ${\bf 4} \otimes {\bf 4}$ of $SU(4)$,
as in the  GS $i$ sector,
in the $(NS-,NS-)$ sector we find
the ${\bf \bar 4} \otimes {\bf \bar 4}$ of $SU(4)$, as in the
GS $-i$ sector \nxtlv.

\ndt Now, the R boundary conditions are:

\eqn\twns{\eqalign{
X^{\mu} ( {\sigma} + 2\pi ) = X^{\mu} (\sigma), & \quad
X^{M} (\sigma + 2\pi ) = - X^{M} ( \sigma)\cr
\psi^{\mu} ( {\sigma} + 2\pi ) = + \psi^{\mu} (\sigma) , & \quad
\psi^{M} (\sigma + 2\pi ) = - \psi^{M} ( \sigma)\cr}}
The zero point energy vanishes and $\psi^{\m}$
have zero modes so that both for the left and right
movers we find spinors in $3+1$ dimensions.
In the $(R+,R+)$ sector the ground state transforms as a product
of two positive chirality Weyl spinors: it is a photon of
positive helicity (this is the state that corresponds to the
ground state of the GS $i$ sector).
In the $(R-,R-)$ sector the ground state
transforms as a product
of two negative chirality Weyl spinors: it is a photon of
negative helicity corresponding to the
ground state of the GS $-i$ sector.

The above construction of the NSR spectrum is identical
to what would be done to describe the ${\IZ}_2$ orbifold of Type
0B theory by reflection of 6 of the coordinates.
Thus, the ${\IZ}_4$ orbifold of Type $\II$B may indeed be thought of as
a ${\IZ}_2$ orbifold of Type 0B.
The bulk states of this theory are simply the
bulk states of Type 0B invariant under the reflection of
6 of the coordinates. In addition, we find a
$3+1$ dimensional massless vector field
which lives on the fixed plane $\IR^{3,1}$.
This effect is somewhat
analogous to what happens in
the $\II$B string on ALE singularity where we find a tensor
theory on the fixed plane $\IR^{5,1}$.

Are there any objects
charged with respect to this vector field?
In the conventional string theory
the quanta of the $U(1)$ field are created by the vertex
operator ${\CV}$ of dimension $(1,1)$.
It can be written in the canonical superghost picture as
follows. Introduce the algebra of the Ashkin-Teller
twist operators
which create the square root cuts for the coordinates 
$X^{i}$ \refs{\dixon,\br}:
\eqn\twst{\eqalign{
{\p} X^{i} (z) {\s}_{j} (0)  \quad = &
\quad {{{\delta}^{i}_{j}}\over z^{1\over 2}}
{\tau}^{j} (0) \cr
{\p} X^{i} (z)  {\tau}^{j} (0) \quad =  & \quad
{{{\delta}^{ij}}\over{\, 2z^{{3\over 2}}\,}} {\s}_{j} (0) +
{{2\delta^{ij}}\over{\,  z^{1\over 2}\,}} {\p \s}_{j}(0) \cr}}
Then
the operator ${\CV}$ is
\eqn\vrtx{{\CV} = {\half} F_{\m \n}(k)
\left( {\g}^{\m} {\g}^{\n} \right)^{\a\b} V_{(-1/2) \a}
{\tilde V}_{(-1/2)\b}} 
where the chiral operator
$V_{(-1/2) {\a}}$ of dimension $1$ is given
by:
\eqn\chrl{V_{(-1/2){\a}}(z) =
e^{-{\varphi}(z) /2} {\Sigma}_{\a} (z)\prod_{M=4}^{9}
{\s}_{M}(z) e^{i k_{\m}X^{\m}(z)} }
with $\varphi$ being the bosonized  superghost and ${\Sigma}_{\a}$ the
product of twist operators for the fermions $\psi^{\m}$ (i.e.
${\Sigma}_{\a}$ is a spinor with respect to the unbroken
part $SO(1,3)$ of the Lorentz group).
The fact that the sum of the dimensions of all the
twist field including that of superghosts is precisely one
is a consistency check of the computation \zrpttw.

We see that it is the field strength, not the gauge field
itself, which enters the vertex operator \vrtx. This is hardly
a surprize given the R-R origin of the $U(1)$ vector
field. Therefore we expect that the only way the
$U(1)$ field may couple to massless open string states
on D-branes is through some kind of a magnetic
moment interaction. On the other hand, it is well-known 
that the D-branes couple
to the R-R fields as the elementary charges do.\foot{
  The point is \tasi\  that in computing the one-point
  disc amplitude the R-R vertex operator
  must be written in the $(-2)$ ghost picture
  while \vrtx\ has only $(-1)$ unit of the total ghost charge.
Upon acting on the operator
  analogous to \vrtx\ by the inverse of the picture
  changing operator
  $e^{\varphi}G_{0}$ one maps the field strength into the
  gauge potential.}
Can we find the
charged states using D-branes?

First of all,
near the fixed plane in the presence
of the large number of the D3-branes
the geometry looks like
$AdS_5 \times {\IR\IP}^5$.
Since the $U(1)$ field comes from the twisted $(R,R)$ sector,
the natural candidates for the charged particles
are the `fractional' D-branes wrapped over 
various cycles of $\IR \IP_5$.\foot{We are grateful to M. Douglas
for suggesting this to us.}
We hope to return to this issue in the future.

\newsec{Open string theory}

Consider the setup of the previous section and
add $N$ $D3$-branes parallel to the fixed fourplane.

\subsec{Gauge theory on the branes in more detail}

It is a standard
by now excersize using the techniques of \DM\ to work out
the field content of the low-energy effective
field theory on the stack of $D3$-branes. Following
\LNV\ we get the fields of the
four dimensional $U(4N)$ ${\CN}=4$ gauge theory
subject to a certain invariance condition:
\eqn\proji{h \cdot \Psi^{Ig \vert J g^{\prime}} =
\Psi^{I gh^{-1} \vert J g^{\prime} h^{-1}}}
where $\Psi$ denotes any field of the $U(4N)$ theory,
$h, g, g^{\prime} \in {{\IZ}}_4$, the pair $Ig$, $I=1,\ldots, N$
is a way to enumerate all $4N$ color indices of the
$U(4N)$ theory, finally $h \cdot \Psi$
denotes the field $\Psi$ transformed by an element $h$ of the
R-symmetry group $SU(4)$.
In particular,
the gauge gluons $A_{\m}$ are singlets
under $h = i^{k}, k = 0,1,2,3$: $h \cdot A_{\m} = A_{\m}$,
the fermions
transform as: $h \cdot \psi_{\a}^{m} = i^k \psi_{\a}^{m}$,
the scalars transform as: $h \cdot \phi^{mn} = i^{2k} \phi^{mn}$.
The solution of the equations \proji\ yields the following
theory: the gauge group is $G = \times_{{\bf i}=0}^{3} U(N_{\bf i})$,
$N_{\bf i} = N$, the global symmetry group is $H = SU(4)$,
the fermions transform in
$({\overline {\bf N}}_{{\bf i}+1}, {\bf N}_{\bf i})
\otimes {\bf 4}$ of $G \times
H$, the scalars are in $({\overline {\bf N}}_{{\bf i}+2},
{\bf N}_{{\bf i}})
\otimes {\bf 6}$, ${\bf i} \equiv {\bf i} +4$.

The scalars have a tree
level potential $V({\phi})$.  Let $\phi_{\bf i}^{M}$ denote the scalar in
$({\overline{\bf N}}_{{\bf i}+2}, {\bf N}_{\bf i})$, $M \in {\bf 6}$ of $H$.
The scalars obey a reality condition: $\phi_{\bf i}^{M, \dagger}
= \phi^{M}_{{\bf i}+2}$.
The potential has the form:  \eqn\scpt{V({\phi}) = \sum_{M < N} {\Tr} \left(
{\phi}_{1}^{M} {\phi}_{3}^{N}  - {\phi}_{1}^{M} {\phi}_{3}^{N} \right)^2 +
{\Tr} \left( {\phi}_{2}^{M} {\phi}^{N}_{0}  -
{\phi}_{2}^{N}{\phi}_{0}^{M}\right)^2} In addition, there are Yukawa
couplings.  Let $\psi_{\bf i}^{m}$ denote the fermion in $({\overline{\bf
N}}_{{\bf i}+1}, {\bf N}_{\bf i})$, $m \in {\bf 4}$ of $H$. Identify ${\bf
6}$ of $H$ with ${\Lambda}^{2}{\bf 4}$:  $M \leftrightarrow [mn] = - [nm]$.
Then the yukawas are:  \eqn\ykws{Y = \sum_{m,n, k,l} {\Tr} \phi_{2}^{kl}
\left( {\ve}_{klmn} \psi_{1}^{m} \psi^{n}_{0} -
\bar\psi_{3}^{\lb} \bar\psi_{2}^{\kb}
\right) + {\rm cyclic}}

We shall now identify these fields with the specific
sectors in the open string theory keeping in mind
further discussion of the interactions of the bulk modes
with the modes propagating on the branes.

\subsec{Vertex operators in the NSR formalism}

We work in the NSR formalism and start with D3-branes placed near,
but not exactly at the fixed
plane.
The plane at ${\IR}^{1,3} \times {\IR}^6/{{\IZ}}_{2}$ parallel
to $\IR^{1,3}$ lifts to two planes on the covering space,
one at $X^M = x^M$, another at $X^M = - x^M$. In addition
we may have electric and magnetic types of branes.
If we start with the configuration of the $N$ self-dual
branes,
then they lift to the $4N$ branes on the covering space, $N$ electric
and $N$ magnetic branes at $X^M = x^M$ and $N$ electric
and $N$ magnetic branes at $X^M = - x^M$.
Hence we get four types of the endpoints for the open strings:
$(\pm x^M)$, $\bf e$ or $\bf m$. From now on we assume that $x^M = 0$.

Let $I,J=1, \ldots, N$ denote the Chan-Paton indices.
The total of $4N$ labels are organized in pairs $(Ig)$,
$I = 1, \ldots, N$, $g = 1, i, -1, -i$.
In {\ttb} the states corresponding to the vector bosons
$W^{Ig \vert J g^{\prime}}$
propagating with the (matrix) polarization vector
$A_{\m}^{Ig \vert J g^{\prime}}(k)$
and the momentum $k_{\m}$ along the $4N$
coinciding D-branes are created by the boundary
vertex operators which can be
written in both $(-1)$ and $0$ pictures:
\eqn\bndry{\eqalign{ V_{-1} (A, k)
\quad = & \quad e^{-\varphi} \left( A \cdot \psi \right) e^{i k
\cdot X} \cr
V_{0} (A, k) \quad = & \quad \left( A \cdot {\p X} +
i ( k \cdot \psi )
(A \cdot \psi)
\right) e^{i k \cdot X} \cr}}
(recall that $A$'s are matrices!) One can also write
down the formulae for the scalar modes.
Using the doubling trick
(see e.g. \KHrev) one gets the same
formulae as \bndry\ except that now the polarization
matrix (denoted by $\phi$) points in the $M$-direction.
Finally, we need the formulae for the vertex operators
describing the emission  of the spinor states with the
wavefunction ${\l}_{Ig \vert Jg^{\prime}}^{\a m}$:
\eqn\bndrys{\eqalign{ V_{-1} ({\l} , k)
\quad = & \quad e^{-\varphi/2} \left( {\l}^{\a m}
\Sigma_{\a} \Sigma_{m} \right)
e^{i k \cdot X} \cr V_{0} ({\l}, k)_{Ig \vert J g^{\prime}}
\quad = & \quad e^{\varphi/2} {\l}^{\b n} {\Gamma}_{i, \b n}^{\a m}
{\Sigma}_{\a}{\Sigma}_{m} \left( {\p
 X}^{i} + i ( k \cdot \psi )  \psi^{i}
\right) e^{i k \cdot X} \cr}}

What remains is to impose the ${{\IZ}}_{4}$ invariance
conditions.  It amounts to demanding:
\eqn\invrc{\eqalign{
A_{\m, Ig \vert J g^{\prime}} \quad = & \quad
\sum_{{\bf i}=0}^{3} A^{IJ}_{\m, {\bf i}}
\delta_{g, g^{\prime}} g^{\bf i} \cr
{\l}_{\a m, Ig \vert J g^{\prime}} \quad = & \quad
\sum_{{\bf i}=0}^{3} {\l}^{IJ}_{\m, {\bf i}, {\bf i}+1}
\delta_{g, i g^{\prime}} g^{\bf i} \cr
{\phi}_{M, Ig \vert J g^{\prime}} \quad = & \quad
\sum_{{\bf i}=0}^{3} {\phi}^{IJ}_{M, {\bf i}, {\bf i}+2}
\delta_{g, - g^{\prime}} g^{\bf i} \cr}}
where the subscripts ${\bf i}, {\bf i}+2$ etc. indicate
the labels of the gauge factors under which the Chan-Paton
factors $I,J$ respectively transform.

We see that in considering the coupling of the
twisted sector bulk mode to the open string states
propagating
on the brane we don't need to
alter the $(-\half, -\half)$ picture of the
vertex operator \vrtx. Hence the only coupling
we might see would be of the magnetic moment type,
i.e. with the field strength of the $U(1)$ field
on the fixed plane.

\subsec{Going away from the singularity}

Let us look at the (classical) vacua of our theory.
The minima of the potential \scpt\ are the solutions
to the equations:
\eqn\vca{
\phi_{{\bf i}}^{M} \phi_{{\bf i} + 2}^{N} =
\phi^{N}_{{\bf i}} \phi_{{\bf i} + 2}^{M}}
By separating the electric stack
from the magnetic stack
we make all the bifundamental fields massive. 
This separation corresponds to the solution
of \vca\ of the form:
$$
{\Phi}^{MN}_{0} \equiv \phi_{0}^{M} \phi^{N,\dagger}_{0} =
x^{M}x^{N} {\bf 1}, \quad
{\Phi}^{MN}_{1} \equiv \phi_{1}^{M} \phi^{N,\dagger}_{1} =
y^{M}y^{N} {\bf 1}
$$
where $x^{M} \sim - x^{M} \neq 0, y^{M} \sim - y^{M} \neq 0$
are the coordinates of the separated stacks.\foot{A more general
configuration is to also separate the branes within each stack
so that the bifundamental fields get unequal masses and
the gauge group is broken down to
the maximal Cartan subgroup.
This configuration corresponds to the diagonal
matrices ${\Phi}^{MN}_{0,1}$ with unequal entries.
For such a configuration we do not
find a non-abelian theory in the infrared.
It is a subtle question
whether the dynamics favors complete
separation of the branes \Zar.}
This vev breaks the gauge group down to the
$U(N) \times U(N)$ subgroup, where $U(N)$'s are the diagonal
subgroups in the $U(N_{0}) \times U(N_{2})$ and
$U(N_{1}) \times U(N_{3})$ respectively.
The fields $\phi_0^M= \phi_2^M$ and $\phi_1^N = \phi_3^N$
act as the six massless adjoint scalars of the first and second
$U(N)$ respectively. Thus, as expected, the $\IZ_4$ theory
Higgses down to the theory on the electric and magnetic branes
of Type 0B theory at a smooth point (away from the singularity).
Indeed, in \KT\ a stack of electric D3-branes was considered
and it was shown that its world volume is described by $SU(N)$
gauge theory coupled to 6 adjoint scalars. The same theory describes
a magnetic stack.

\newsec{Discussion}

In this paper we have considered an orbifold of Type $\II$B string theory
by reflection of six of the coordinates.
Because of the action on the fermions
this is actually a $\IZ_4$ orbifold of Type $\II$B which can equivalently
be thought of as a $\IZ_2$ orbifold of Type 0B string theory.
Consideration of the closed strings in the twisted sector led us
to conclude that the only massless field
on the fixed fourplane $\IR^{3,1}$ is a photon.
Thus, we have found a simple way of embedding a $U(1)$
gauge theory into string theory.

In view of recent progress on connections between large $N$ gauge theories
and strings, it is even more interesting to find an embedding of pure
glue $U(N)$ gauge theory into a string theory which might help in searching
for its stringy dual. The pure glue $U(N)$ gauge theory may be
constructed as a $\IZ_4$ orbifold of the ${\cal N}=4 $ supersymmetric
$U(N)$ gauge theory by the center of the $SU(4)$ R-symmetry which
acts trivially on the group indices. This is known as an `irregular'
orbifold theory because the gauge group twists do not satisfy
the condition $\Tr \gamma = 0$. If we think of the 4 basic `fractional'
D3-branes of Type $\II$B on $\IR^6/\IZ_2$ as corresponding to the 4
different nodes of the quiver diagram, then the pure glue theory
results on a stack of $N$ `fractional' D3-branes of the same type.
The absence of scalar fields in this theory is related to the fact
that the fractional branes can exist only on the fixed fourplane
and hence cannot fluctuate away from it.
Such a stack has non-vanishing tadpoles for the twisted states, 
and in particular
for the tachyon. Thus, as for the stack of electric D3-branes, the dual
gravity background will have a radially varying tachyon field 
\refs{\KT,\JM,\KTnew}. We
postpone a study of this background for the future.

We have also initiated a study of interactions between the photons on
the fixed fourplane with the gauge theories on the D3-branes.
Overall, this orbifold is quite different from the
${\cal N}=2$ $\IZ_2$ orbifold studied in \DM. In that case the
scalars on the fixed sixplane could be used to blow up the
$\IR^4/\IZ_2$ into a smooth ALE space. For $\IR^6/\IZ_2$ we find no
scalars in the twisted sector, hence it is not clear how to resolve
the singularity.

The original motivation for our work was to find a D-brane realization
of the orbifold of ${\cal N}=4$ SYM by the $\IZ_4$ center of its $SU(4)$
R-symmetry. We have found such a realization in terms of D3-branes
of Type 0B string theory orbifolded by reflection of six coordinates.
In general, it is quite clear that if the orbifold group
$\Gamma$ is a subgroup of
$SO(6)$ then the gauge theory may be realized on D3-branes of
Type $\II$B theory,  but if $\Gamma$ also includes the center of
$SU(4)$ then the realization involves the D3-branes of Type 0B theory.
\bigskip
\bigskip
\appendix{A}{Another ${\IZ}_4$ orbifold of Type $\II$B string}

It may be interesting to consider orbifolding by reflection of two
of the coordinates:
\eqn\newrefl{
X^{M} \to - X^{M}, \quad M = 8, 9\ .
}
This case is similar to the reflection of 6 coordinates
in that it is a
${\IZ}_4$ operation on {\ttb}.
Similarly, it can be thought
of as a ${\IZ}_2$ orbifold of Type 0B theory. This theory has
an $SO(2)$ internal symmetry group and $SO(6)$ transverse group:
so compared to the orbifold \refl\ the two are
interchanged.

The discussion of the spectrum proceeds analogously to the case
\refl.
The NSR untwisted sector
gives states of Type 0B theory invariant under
\newrefl. Something interesting happens in the twisted sector
which is localized on the fixed eightplane $\IR^{7,1}$.
The $(NS,NS)$ zero-point energy is
$$ E_L= E_R=- 1/4\ , $$
so it seems that there is a tachyon living on the fixed plane.
However, the ground state transforms as a vector
  $A_M$ under the
$SO(2)$ rotation, hence it is projected out!

Thus, the only massless twisted sector states come from the RR sector.
{}From the $(R+,R+)$ sector we find
${\bf 4} \otimes {\bf 4}$ of $SU(4)$,
which gives a 1-form and a selfdual
3-form of $SO(6)$. From the
$(R-,R-)$ sector we find
${\bf \bar 4} \otimes {\bf \bar 4}$ of $SU(4)$,
which gives a 1-form and an anti-selfdual
3-form of $SO(6)$. Putting them together we thus find
the gauge potentials, $A_\mu$, $B_\mu$
  and $A_{\mu\nu\lambda}$
which correspond to
  two 2-form field strengths and one 4-form field
strength.
Dualizing one of the 2-forms we have
$F_2$, $F_4$ and $F_6$. These naturally couple to a
  0-brane, a 2-brane
and a 4-brane in 8 dimensions. It seems plausible
  that these objects come
from
1-branes, 3-branes and 5-branes wrapped over the
${\IR\IP}^1
 \approx {\bf S}^1$
  which is the base
of the cone. This
8-dimensional theory deserves further study.

\bigskip
\bigskip
\noindent
{\bf Acknowledgements}
\bigskip
We are grateful to O.~Bergman, M.~Douglas, A.~Losev and L.~Paniak
for useful discussions.
The work of I.~R.~K. was supported in part by the NSF
grant PHY-9802484 and by the James S. McDonnell
Foundation Grant No. 91-48;
that of N.~A.~N by Harvard Society of Fellows,
  partly by NSF under the
grants
PHY94-07194,
  PHY-98-02709, partly by
RFFI under grant 98-01-00327,
partly  by the grant
96-15-96455 for scientific schools. Research of
S.~L.~Sh. is supported
 by DOE grant
DE-FG02-92ER40704, by NSF CAREER award,
  by OJI award from DOE and by Alfred
P.~Sloan foundation. N.~A.~N. is grateful to ITP, Santa Barbara
and to ESI, Vienna  
for hospitality while some parts of this work were being done.


\vfill\eject
\listrefs
\end